\documentclass[12pt]{iopart}

\usepackage{graphicx}

\begin{document}

\title[Tunable few electron quantum dots in InAs nanowires]{Tunable few electron quantum dots in InAs nanowires}

\author{I. Shorubalko, A. Pfund, R. Leturcq, M. T. Borgstr\"om, F. Gramm, E. M\"uller, E. Gini and K. Ensslin}

\address{Solid State Physics Laboratory, ETH Zurich, 8093 Zurich, Switzerland}

\ead{ensslin@phys.ethz.ch}

\begin{abstract}
Quantum dots realized in InAs are versatile systems to study the effect of spin-orbit interaction on the spin coherence, as well as the possibility to manipulate single spins using an electric field. We present transport measurements on quantum dots realized in InAs nanowires. Lithographically defined top-gates are used to locally deplete the nanowire and to form tunneling barriers. By using three gates, we can form either single quantum dots, or two quantum dots in series along the nanowire. Measurements of the stability diagrams for both cases show that this method is suitable for producing high quality quantum dots in InAs.
\end{abstract}

\maketitle

\section{Introduction}

Semiconductor quantum dots (QDs) have been widely studied due to the possibility to manipulate the spin of single electrons, and they have been proposed as building blocks for quantum computing \cite{Loss01}. Recent experiments have demonstrated measurements and manipulations of single spins, as well as coupling of two spins, which are the elementary functions for using QDs as quantum bits \cite{Elzerman03,Petta04}. Interactions with nuclear spins are thought to be the main mechanism for spin decoherence in GaAs \cite{Petta04}. Other materials with a strong spin-orbit interaction could provide an opportunity to study different decoherence mechanisms, such as spin-orbit interaction. In addition, strong spin-orbit interaction could provide the opportunity to manipulate single spins in QDs using an electric field instead of a magnetic field \cite{Stepanenko01,Stepanenko02,Debald01,Flindt01}. InAs is a material with strong spin-orbit interaction, and QDs defined in InAs might give the opportunity to investigate these effects.

Attempts to realize gate-defined QDs for transport experiments in InAs face the problem of producing reliable Schottky barriers, due to the small energy gap of this material \cite{Kajiyama01}. For this reason, the method of top-gates, which is very successfull for realizing tunable nanostructures in GaAs heterostructures, is not straight forward on InAs. It has been shown that QDs can be realized in planar InAs heterostructures \cite{Jones01}. High quality QDs have been realized on nanowires (NWs), either by using Schottky contacts on InP \cite{Franceschi02} and Si NWs\cite{Zhong01}, or by epitaxial growth of InP barriers along an InAs NW \cite{Thelander01,Bjork03}. While these structures could demonstrate quantum Coulomb blockade down to the last electron, one drawback is the difficulty to control independently the coupling of a single QD to its leads or to another nearby QD. More recently, a new technique consisting of depositing an InAs NW on predefined finger gates covered by a silicon nitride isolation layer could demonstrate fully tunable single and double QDs, but was limited to QDs containing a large number of electrons \cite{Fasth01}.

Here we show another method to fabricate QDs along InAs NWs. We use top-gates in order to locally deplete the NW and form the tunnel barriers necessary to observe Coulomb blockade. The metallic gates are electrically isolated from the conducting channel of the NW by the native oxide covering the NW after growth. This way, "half wrap-around-gates" covering a big fraction of the NW's surface are created and they are expected to be more efficient than earlier attempts using local back-gates \cite{Bryllert01}. By measuring the stability diagram in two configurations, one with a single QD, the second with two coupled QDs, we show that this method is very suitable for producing high quality few electrons QDs in InAs for transport experiments.

\section{Samples and experimental methods}

InAs NWs are grown on GaAs $\langle$111$\rangle$ substrates by metal-organic vapor phase epitaxy (MOVPE) \cite{XiaY01}. Au colloids of diameter 20 to 60 nm are deposited on the substrate, and serve as catalyst for the growth. Nanowires are 5 to 10 $\mu$m long and the diameter varies from maximum 200 nm at the base down to less than 50 nm at the top.

Results of two different samples (referred as A and B) will be presented, for which the fabrication processes differ slightly. The NWs are deposited on a highly doped Si wafer covered by 300 nm of SiO$_2$, either after being put into an ethanol solution (sample A), or directly by mechanical transfer (sample B). The highly doped Si substrate is used as a back-gate to modulate the electron density in the whole NW. Ohmic contacts to the NWs are defined using optical lithography, on part of a NW of diameter 150 nm (sample A) or 100 nm (sample B). After development, the contact areas are passivated using (NH$_4$)$_2$S$_x$ \cite{Oigawa01}, and layers of Ti (thickness 20 nm) and Au (180 nm) are evaporated. This method gives a contact resistance below 100 $\Omega$.

\begin{figure}
\includegraphics[width=.9\textwidth]{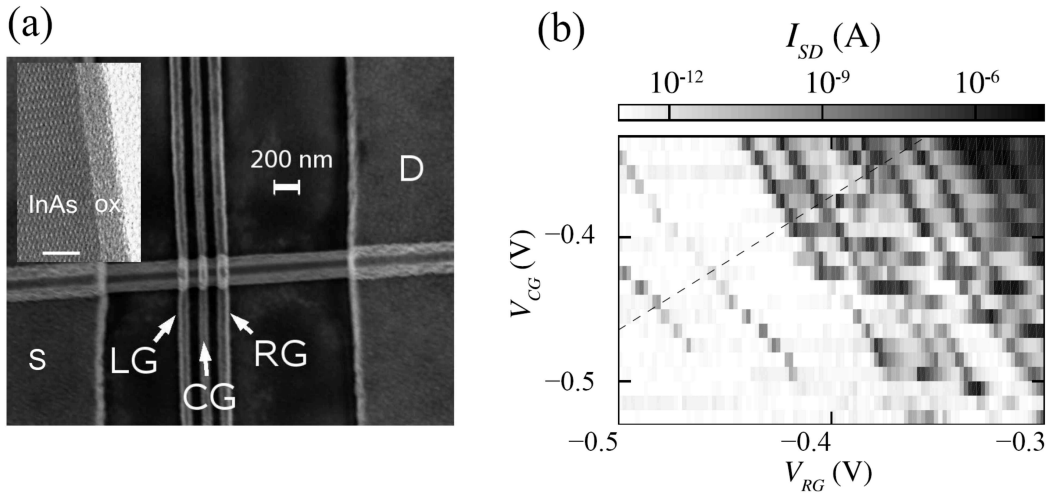}
\caption{(a) Scanning electron micrograph of the sample A, showing the ohmic contacts labelled S and D, and three finger top gates labelled LG, CG and RG. Inset: transmission electron microscope image of the cross-section of an InAs nanowire, showing the native oxide layer. The scale bar is 3 nm. (b) Current through a single QD formed in sample A at 1.7 K as a function of the voltage applied on the right gate, $V_{RG}$, and the voltage applied on the center gate, $V_{CG}$. The back gate voltage is +6 V, and the voltage of the left gate is kept to +0.7 V in order to keep the lead open. Each line correspond to the addition of an electron in the QD. The slope of the lines is -0.6, i.e. of the order of -1, as expected for a QD formed inbetween the gates RG and CG. The dashed line is the line followed to take the charge stability diagram of Fig.~\ref{fig2}.}\label{fig1}
\end{figure}

In order to realize tunneling barriers along the NWs, top gate fingers are defined using electron beam lithography, and evaporation of a double layer of Cr and Au (with thicknesses of respectively 10 and 100 nm for sample A, 6 and 66 nm for sample B respectively). No surface preparation is done before the metal evaporation, in order to keep the native oxide of about 1.5 to 2 nm thickness on InAs (see inset of Fig.~\ref{fig1}). This thickness is comparable to the native oxide measured on InAs layers \cite{Hollinger01}. This way we obtain a capacitive coupling of these metallic stripe and the NWs. Such a homomorphic insulating layer has been shown to produce low quality macroscopic metal-insulator-semiconductor structures on laterally extended InAs structures \cite{Wieder01}. However, we show here that on the length scale of 100 nm, the metallic gates do not show breakdown for voltages in the range $\pm 1$V and are appropriate in order to locally deplete the NW.

A scanning electron microscope image of sample A is shown in Fig.~\ref{fig1}(a). The three finger gates have a width of approximately 80 nm and a periodicity of 150 nm. Sample B has the same geometry, with a width of 60 nm and a periodicity of 120 nm, evaluated from the lithographic design. The lower limit for these widths is given by the thickness of the evaporated metal layers, since a width-height ratio of at least 1:1 is required for our lithographic structures. Samples A and B have been measured respectively in a pumped $^4$He system with base temperature 1.7 K, and a $^3$He/$^4$He dilution refrigerator with base temperature 30 mK.

\section{Single quantum dot}

With all gates put to ground, transport measurements on sample A at 1.7 K show a high resistance, due to the complete pinch-off of the NW below the top gates. This depletion due to the presence of the gates is not observe with the same magnitude for all devices, and could be due to charges trapped in the oxide layer between the metallic gates and the NW. The original resistance of the NW is recovered by either applying a large positive voltage on the back gate or on the top gates.

In order to form a single QD, defined by two tunnel barriers, two gates (RG and CG) are set to a low potential while the third gate (LG) is kept at +0.7 V. In this regime, the two-point conductance measured between source and drain shows peaks when sweeping right and center gates, which is a clear signature of Coulomb blockade. These peaks correspond to the alignment of the discrete levels in the QD with the chemical potential of the leads, which allow transport of electrons. The current is blocked in-between the peaks due to Coulomb repulsion between the electrons \cite{Kouw01}. To show that the QD is created really between both gates RG and CG, the Coulomb blockade peaks are measured when varying the voltages $V_{RG}$ and $V_{CG}$, as shown in Fig.~\ref{fig1}(b). The slopes of the Coulomb peak lines are of the order of -1, showing that the lever arms of both gates on the QD are comparable. This proves that the QD is situated between the gates. We have checked that using the gates LG and CG, and keeping RG to a large voltage, it is also possible to form a QD between these two gates. We are therefore able to create independently two dots along the NW using the three gates.

\begin{figure}
\includegraphics[width=.9\textwidth]{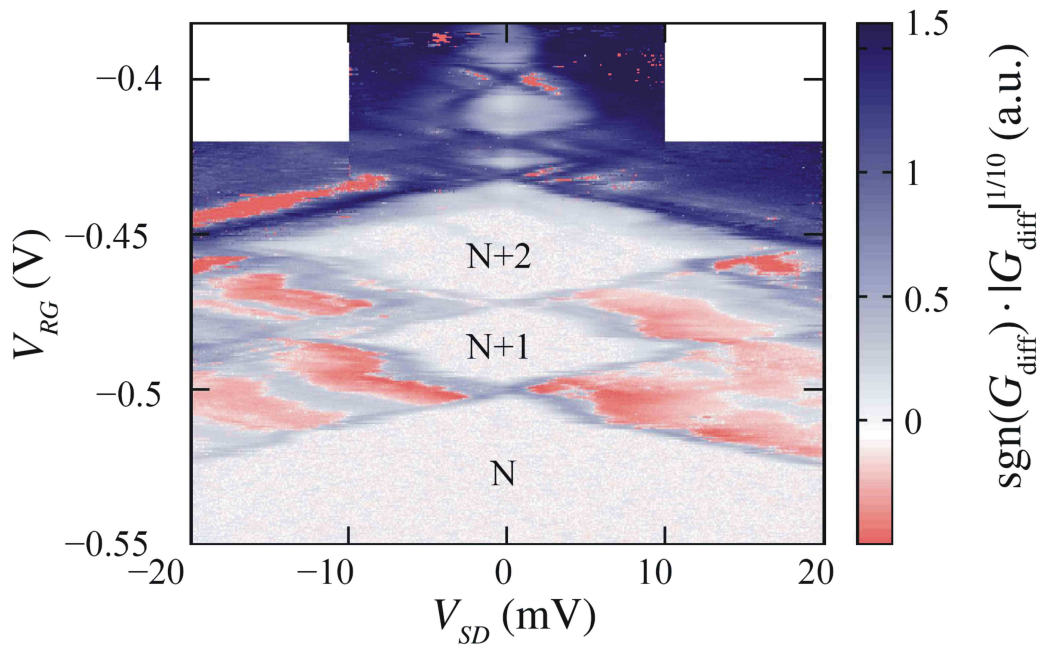}
\caption{Differential conductance $G_{diff}=dI_{SD}/dV_{DS}$ as a function of the bias voltage $V_{DS}$ and the gate voltage $V_{RG}$ measured on sample A at 1.7 K. In this measurement, the gate voltage $V_{CG}$ is changed proportionally to $V_{RG}$, $V_{CG}= 0.99 V_{RG} + 0.042$ V.}\label{fig2}
\end{figure}

The properties of the QD can be determined by measuring the conductance at large bias voltage. Figure~\ref{fig2} shows the charge stability diagram of sample A, measured by sweeping the source-drain bias voltage, and changing the global potential of the dot symmetrically with the gates RG and CG as indicated by the line in Fig.~\ref{fig1}(b). In this picture, the white regions correspond to blockaded transport. A finite conductance can be observed either by aligning the potential of the QD with the chemical potential of the leads, or by applying a bias voltage large enough in order to compensate the charging energy of the QD. From the size of the first diamond (labelled $N$ in Fig.~\ref{fig2}) we determine the charging energy of this QD, $E_C \approx 11.5$ meV. This energy corresponds to the electrostatic energy of a capacitor $C=13.9$ aF, which is the capacitance of an isolated sphere of radius $R = C/(4 \pi \epsilon_0 \epsilon_r) = 8.3$ nm ($\epsilon_r = 15$ for InAs).

The large difference between the first and second diamond (from the bottom of the diagram, corresponding to $N+1$ and $N+2$ electrons in the QD) is attributed to the quantum confinement in a spin-degenerate system. The effective mass of conduction electrons in InAs is rather small, $m_{eff} \approx 0.02 m_0$, with $m_0$ the electron mass. Electron-electron interaction effects are therefore expected to be smaller than in GaAs-based quantum dots. It is therefore very likely that the orbital states are successively filled by spin-up and spin-down electrons. Starting with $N$ even, an additional $(N+1)^{th}$ electron needs to pay only the charging energy $E_C$ since it can occupy the same orbital state as the $N^{th}$ electron, with an opposite spin. Because of the Pauli principle, the $(N+2)^{th}$ electron needs to pay an energy equals to $E_C + \Delta_{i,i+1}$, where $\Delta_{i,i+1}$ is the quantum level spacing between the orbital level $i$ occupied by the $N^{th}$ electron and the orbital level $i$ occupied by the $(N+2)^{th}$ electron. From this difference of energy we determine $\Delta \approx 7.5$ meV. Compared to the confinement energy of a 3-dimensional quantum well, $\Delta = \hbar^2 \pi^2 /(16 m^* R^2)$, this energy corresponds to a radius $R=17$ nm, which is close to the size determined from the charging energy.

Figure~\ref{fig2} is qualitatively very similar to charge stability diagrams of QDs which can be depleted down to the last electron \cite{Kouwenhoven05}.  In our system, we are, however, not able to determine whether $N=0$ for the lowest Coulomb blockaded region, since applying a more negative gate voltage on RG and CG will also close the tunnel barriers from source and drain, which suppresses the current through the QD. In addition, from the size determined above and the approximate electron density of $10^{18}$ cm$^{-3}$ determined in the ungated wire, we evaluate the number of electron to be less than 10.

\begin{figure}
\includegraphics[width=.9\textwidth]{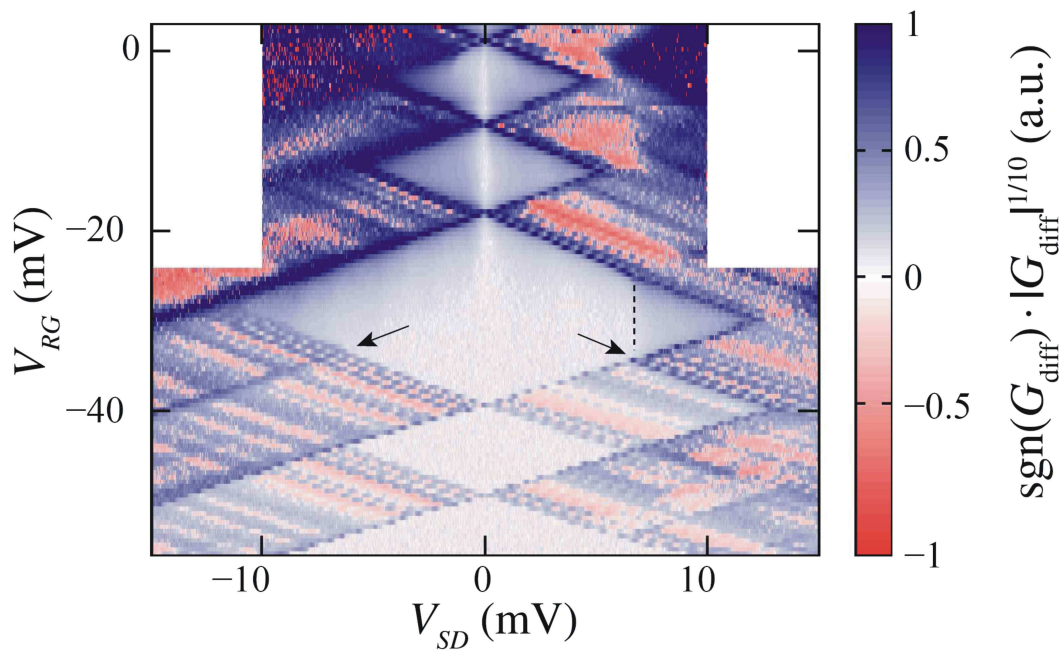}
\caption{Differential conductance $G_{diff}=dI_{SD}/dV_{DS}$ as a function of the bias voltage $V_{DS}$ and the gate voltage $V_{RG}$ measured on sample B at 30 mK. In this measurement, the gate voltage $V_{CG}$ is also changed proportionally to $V_{RG}$, $V_{CG}= 0.85 V_{RG} -0.124$ V. The back gate voltage is set to 0 V. The arrows point on lines corresponding to excited states, and the dashed line emphasizes the transition to the inelastic cotunneling. The electronic temperature is 150 mK, determined from the width of the Coulomb peaks at low bias voltage \cite{Kouw01}.}\label{fig3}
\end{figure}

In the measurement done at 1.7 K, the quantum confinement is not clearly seen in the excitation spectrum at high bias voltage, probably due to the time dependent fluctuations observed in the Coulomb diamonds. These presumably stem from an insufficient screening of the experiment with respect to external noise in this particular cryostat, which can induce fluctuations in the confinement potential of the QD. To show that these fluctuations are due to the set-up, and are not intrinsic to the device, we have performed a measurement at lower temperature with a system specially designed to filter external noise. The measurement of the Coulomb diamonds on sample B at an electron temperature of $145 \pm 10$ mK, determined from the Coulomb blockade peak width \cite{Kouw01}, is displayed in Fig.~\ref{fig3}. This measurement shows a much more stable diagram, and lines corresponding to excited states are clearly seen (arrows in Fig.~\ref{fig3}). In addition, we can also resolve a step in the Coulomb blockade region due to inelastic cotunneling through the corresponding excited state (dashed line in Fig.~\ref{fig3}) \cite{Franceschi03}.

\section{Double quantum dot}

\begin{figure}
\includegraphics[width=.9\textwidth]{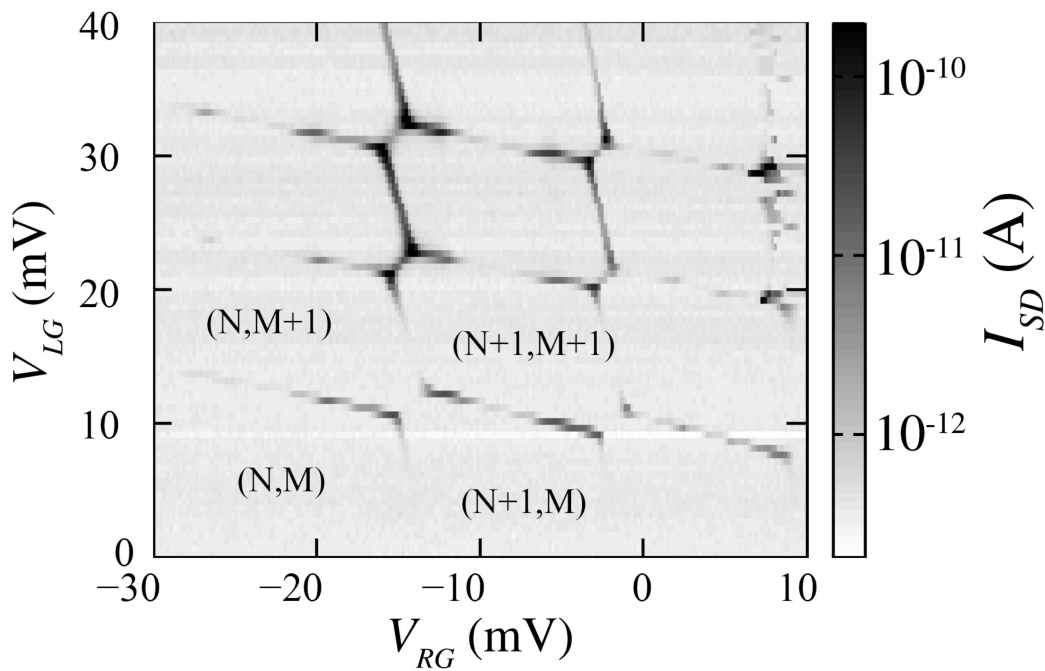}
\caption{Current measured through the double QD defined in sample B at 30 mK, as a function of both gate voltage $V_{RG}$ and $V_{LG}$. The voltage $V_{CG}$ is kept constant at -120 mV, and the back gate voltage is set to 0 V. The number of charges in the right QD is indicated by $N$, $N+1$..., and the number of charges in the left QD by $M$, $M+1$,....}\label{fig4}
\end{figure}

Using now the three gates to form three tunnel barriers, we show that we can form two coupled QDs. Figure~\ref{fig4} shows the charge stability diagram of the double QD, while changing the voltages on gates RG and LG in order to tune the number of electrons in each QDs independently. The voltage on CG is kept at -120 mV. In the resulting measurement, vertical lines correspond to charging of the right QD (with number of electrons indicated by $N$, $N+1$,...), and the horizontal lines to the charging of the left QD (with number of electrons indicated by $M$, $M+1$,...). The hexagon pattern obtained is characteristic of a double dot with strong capacitive coupling between both dots \cite{vdWiel01}. Beyond that, we observe a finite current in the spacing between the two degeneracy points corresponding to electron and hole transport through the DQD respectively. This signature of strong tunnel coupling between both QDs is required in order to form molecular states in this system \cite{Blick01}.

\section{Conclusion}

We have realized few electron quantum dots in InAs nanowires using top gate fingers in order to create tunneling barriers along the nanowire. This method allows to obtain fully tunable quantum dots as well as double quantum dots coupled in series. We demonstrate that these devices are perfectly suitable to study single electron transport, as well as coupled systems in the few electron regime.

\section*{Acknowledgements}
We thank ETH Zurich for financial support. IS thanks European Commission for a Marie-Curie fellowship.


\end{document}